\documentclass{pnastwo}

\usepackage{titlesec} %% for squeezing

\usepackage{graphicx}
\usepackage{amssymb,amsfonts,amsmath}
\usepackage{dcolumn}
\usepackage{booktabs}

\begin{document}
	%------------------------------------------------
	% FOR SQUEEZING PURPOSES
	%------------------------------------------------
	
%	\titlespacing\section{4pt}{0pt plus 4pt minus 2pt}{0pt plus 2pt minus 2pt}
%	\titlespacing\subsection{4pt}{0pt plus 4pt minus 2pt}{0pt plus 2pt minus 2pt}
%	\titlespacing\subsubsection{4pt}{0pt plus 4pt minus 2pt}{0pt plus 2pt minus 2pt}
%	\setlength{\belowdisplayskip}{4pt} \setlength{\belowdisplayshortskip}{4pt}
%	\setlength{\abovedisplayskip}{4pt} \setlength{\abovedisplayshortskip}{4pt}

	%---------------------------------------------------------------
	%---------------------------------------------------------------
%\setlength{\parindent}{15pt}
%\renewcommand{\baselinestretch}{1.5}
%\setlength{\parskip}{20pt}

	\title{Ribosome self-assembly leads to overlapping reproduction cycles and increases growth rate}	
	%\title{Overlapping ribosome reproduction cycles via a self-assembly step increase growth rate}	
	
	\author{
		Rami Pugatch
		\affil{1}{Department of Industrial Engineering and Management, Ben-Gurion University of the Negev,
Beer Sheva, 84105, Israel} 
		\affil{2}{Quantitative Life Science Section, The Abdus Salam International Center for Theoretical Physics, Strada Costiera 11, 34014, Trieste, Italy},
		Yinon M. Bar-On
		\affil{3}{department of Plant and Environmental Science, Weizmann Institute of Science, Rehovot 7610001, Israel} 
		}%

%	\contributor{Submitted to Proceedings of the National Academy of Sciences
%		of the United States of America} 

	\significancetext{The transcription-translation machinery has a dual role to synthesize the proteome, and to synthesize copies of itself. Two key players in this process, RNA-polymerases and ribosomes, self-replicate by jointly producing their sub-components which subsequently self-assemble to new RNA-polymerases and ribosomes. We show that a self-assembly step allows ribosomes to perform more tasks, including starting another round of self-reproduction, prior to the completion of the previous round. This overlapping of self-reproduction cycles increases growth rate relative to the serial case. We devise a model for concurrent self-replication with a self-assembly step and employ it to infer in-vivo duration of ribosome self-assembly in fast growing \textit{E. coli} and predict the effect of limiting assembly chaperons on the growth rate.}
	
\maketitle
\begin{article}
\begin{abstract}{In permissive environments, \textit{E. coli} can double its dry mass every $\sim 21$ minutes. During this time, ribosomes, RNA polymerases, and the proteome are all doubled. Yet, the question of how to relate bacterial doubling time to other biologically relevant time scales in the growth process remains illusive, due to the complex temporal organization of these processes. In particular, the relation between the cell's doubling time and the ribosome assembly time is not known. Here we develop a model that connects growth rate to ribosome assembly time and show that the existence of a concurrent ribosome self-assembly step increases the growth rate, because during ribosome self-assembly existing ribosomes can start a new round of reproduction, by making a new batch of ribosomal proteins prior to the completion of the previous round. This overlapping of ribosome reproduction cycles increases growth rate beyond the serial-limit that is typically assumed to hold. Using recent data from ribosome profiling and established measurements of the average translation rate, rigid bounds on the in-vivo ribosome self-assembly time are set, which are robust to the assumptions regarding the biological noises involved. Utilizing these physiological parameters, we find that at $21$ minutes doubling time, the ribosome assembly time is $\sim 6$ minutes --- three fold larger than the common estimate. We further use our model to explain the detrimental effect of a recently discovered ribosome assembly inhibitor drug, and predict the effect of limiting the expression of ribosome assembly chaperons on the overall growth rate.}
		\end{abstract}
		%Abstract. Provide an abstract of no more than 250 words on page 2 of the manuscript. Abstracts should explain to the %general reader the major contributions of the article. References in the abstract must be cited in full within the abstract %itself and cited in the text.
		\keywords{self-assembly | ribosome | growth rate | branching processes}
		%\abbreviations{RNAP, ; DNAP, }

%-----------------------------------------------0-
% INTRODUCTION begins 
%------------------------------------------------
		\dropcap{A}ll known single-cell organisms share the same basic architecture first suggested by Von-Neumann \cite{VN1,ROPP}. In particular, all single cells have a membrane, metabolic machinery that is responsible for supplying ample amounts of energy and substrates, a transcription-translation machinery, and DNA to instruct it. The cell reproduces by allowing the transcription-translation machinery to read the DNA instructions, and make copies of all the molecular machines in the cell, as well as copies of itself. Concurrently to this process the preexisting molecular machines, produced in previous production rounds, keep supplying energy and substrates, produce membrane bound volume and replicate DNA.  
		
Complex production and assembly processes are comprised of many indivisible tasks that are constrained to occur according to a given partial temporal order. This partial temporal order forces some tasks to occur in series, allowing other tasks to occur concurrently. When all tasks are completed, a functional end product emerges. The duration of the longest set of tasks that are bound to occur in series defines a natural time scale, known as the critical path duration, which sets a lower bound on the production time of a specific product. 

There are three generic methods to increase the production rate of such complex processes. A straightforward method to increase the rate is to decrease the critical path duration $T_c$. If this is impossible e.g. due to constraints such as accuracy \cite{SpeedAccuracyTradeoff}, there are two quintessential alternatives that allow production rate to become larger than the reciprocal of the critical path duration: (i) parallel production --- having multiple production lines that run in parallel; (ii) pipelining --- on a single production line, starting a new round of production prior to the completion of the previous round. 

To illustrate these three methods for increasing the production rate, consider a single ribosome translating mRNA. The critical path duration is the average translation time. To get a production rate that is twice as fast, we can double the speed of translation --- reducing by half the critical path duration. Alternatively, we can parallelize the production line, by doubling both the number of mRNA's and the number of translating ribosomes per mRNA. As a result, the rate of protein production will double but the critical path duration will not change.

Finally, we can use pipelining, by allowing ribosomes to start a new round of protein production prior to the completion of the previous round. For example, if on a single strand of mRNA we allow a second ribosome to start a new round of protein translation while the previous ribosome is on average midway in the process of making the previous round, at steady state we will increase the production rate by a factor of two, without reducing the critical path duration. This factor-of-two increase will be valid as long as neighboring translating ribosomes are far enough apart to be effectively decoupled from each other in order to avoid extra delays caused by self-exclusion (``traffic jams''). Having multiple mRNA’s each pipelined with multiple ribosomes allows the methods of parallelization and pipelining to be combined.

Perhaps surprisingly, in the context of self-replication, only two of the three methods help to decrease the doubling time. The argument is simple --- let $U$ be a self-replicating machine which can make a copy of itself in $T_c$ time units, where $T_c$ is the critical path duration for making a single copy. Let's investigate how we can increase the doubling rate. One straight forward way to increase the doubling rate is to decrease the critical path duration $T_c$. What about parallelization? If we start with two machines instead of one, the production rate --- how many U's are made per unit time, will increase by a factor of two. However, the doubling time will remain $T_c$ since now, in order to double, we need to produce two $U$-machines rather than one. 

More generally, if we start with $n$ self-replicating machines that need to double to $2n$ machines, increasing the number of machines at the beginning of the process to $2n$ will not increase the doubling time, since the target will be to reach $4n$ machines. We conclude that parallelization does not help increasing the doubling rate. 

What about pipelining? If the machine $U$ can start the process of making a new copy but can leave this process in the middle in order to restart another round of $U$ production while the immature $U$ machine continues to mature independently, then the growth rate will increase without changing the critical path duration. We call this process pipelined self-replication (\cite{ROPP}, chapter 7). The mathematical structure of such a processes turns out to be identical to a branching model first studied by Crump and Mode and by Jagers \cite{Jagers}, in the context of population demography and also in the study of budding yeast \cite{PrionPaper}, and is known as a general branching process or Crump-Mode-Jagers branching process \cite{Jagers,Kimmel}.

An important recent result in the field of bacterial physiology is the bacterial growth law developed and experimentally tested in \cite{hwa1}. This bacterial growth law connects the growth rate $\mu$, the ribosome workload --- the time it takes a ribosome to translate all ribosomal proteins $\tau_{RP}$, and the percentage of ribosomes allocated toward making ribosomal proteins $\alpha$, $\mu=\frac{\alpha}{\tau_{RP}}$ (see also \cite{hwa2,hwa3,dill2}). 

Here we present a novel quantitative relation (``growth law'') that connects the growth rate $\mu$ to the ribosome assembly time $\tau_{SA}$ as well as other more standard cellular parameters, specified below. The connection between growth rate and assembly time is not trivial, since in the presence a non-zero assembly time, the process of self-reproduction of ribosomes is pipelined, i.e. several generations are produced concurrently, in contrast to the tacitly assumed serial reproduction model, where the overall ribosome doubling time is simply the assembly time plus the time to produce all the ribosomal proteins. 

We develop a model that can robustly predict rigid bounds on the in-vivo ribosome self-assembly duration, and find that the assembly time is roughly three times higher than the prevalent estimate. We further employ our model to explain the relation between the newly discovered effect of the drug Lamotrigie on ribosome self-assembly in live \textit{E. coli} \cite{Lamotrigine} and its growth. Finally, we study dependence of the growth rate on the reliability of the assembly process, which can also be experimentally tuned by limiting assembly chaperons.

\subsection{A new growth law that accounts for ribosome self-assembly and noise.} Our main result is an implicit functional relation (``growth law'') between the fraction of ribosomes busy translating ribosomal proteins --- $\alpha$, the single cell biomass growth rate --- $\mu$ and the following Laplace transforms: (i) Laplace transform of the distributions of ribosome idling times --- $P_0(s)$; Laplace transform of the distribution of durations to translate the $i^{th}$ ribosomal protein --- $P_{i}(s)=P(s)$ assumed equal among all ribosomal proteins \cite{shlomi}; (iii) Laplace transform of the distribution of ribosome assembly times --- $P_{SA}(s)$. The Laplace-transform naturally appears in this problem because upon averaging, a complicated convolution of many random independent factors e.g. assembly and translation durations, factor under its operation thus leading to a simpler equation \cite{Jagers}, (see SI, section III). The equation we obtain is,
\begin{eqnarray}\label{EqOne}
		P_0(s)P(s)(\frac{\alpha}{n} P_{SA}(s) + 1)=1,
\end{eqnarray}
where $n=54$ is the number of ribosomal protein subunits in a ribosome (there are $52$ ribosomal protein species but two of them appears in tandem dimers and hence $n = 54$ \cite{WilliamsonReview}). 
The rate parameter $s$ that solves Eq. \ref{EqOne} above is the asymptotic growth rate, which is the growth rate of a large asynchronous collection of ribosomes, under the assumption that the supply of material inputs and rRNA is not limiting. Our formalism allows us to derive other growth laws, when one of these assumptions breaks (SI, section III). 

To illustrate the use of Eq. \ref{EqOne} consider a deterministic setting with all the process durations fixed. The resulting growth law is then $\frac{\alpha}{n} e^{- \mu (\tau_0 + \frac{\tau_{RP}}{n} + \tau_{SA} )}+e^{- \mu (\tau_0 + \frac{\tau_{RP}}{n} )} =1$ with $\tau_{RP}$ is the ribosome workload --- duration to make all the ribosomal proteins by a single ribosome, $\tau_{SA}$ is the ribosome self-assembly time, and $\tau_0$ is the ribosome rest time between consecutive translations.

\subsection{Two models for ribosome self-assembly.} Before deriving Eq. \ref{EqOne}, we discuss how we model the self-assembly process --- a crucial ingredient in the derivation. Our first assembly model is very simple, assuming that whenever all $54$ ribosomal proteins are present in stoichiometry --- one per type, one ribosome assembly process is initiated which will end after an assembly duration $\tau_{SA}$. We summarize this by the following reaction equation
\begin{eqnarray}\label{SAreactionScheme}
		RP_1+ \ldots + RP_{54} + rRNA \rightarrow R^*,
\end{eqnarray}
where $R^*$ is a new ribosome.

Actual ribosome assembly is significantly more elaborate. As discovered by Nomura \cite{Nomura}, Nierhaus \cite{Nierhaus}, Williamson \cite{Williamson1} and others \cite{WilliamsonReview} the process of ribosome assembly proceed according to a partial temporal order. Some proteins cannot bind before other proteins are docked and after the sub-assembled ribosome had properly conformed (also see SI, section II). 

To test the sensitivity of our growth law to the intricacies of the self-assembly process we also considered a second model, where we roughly arranged all the ribosomal proteins into three groups. The first group is the primary binders --- ribosomal proteins that directly attach to rRNA, if it is present.  After a duration $\tau_{SA_1}$, the sub-assembled ribosome we denote by $A$ is ready and can allow the secondary binders --- the second group of ribosomal proteins, to attach to it. After a duration $\tau_{SA_2}$ the sub-assembled structure we denote by $B$ is formed, and the third group of ribosomal proteins can bind to it thus forming, after a duration $\tau_{SA_3}$ a new ribosome. We summarize the second model as:
\begin{eqnarray}\label{SAreactionScheme2}
		 RP_1+ \ldots + RP_{l_1} + rRNA \rightarrow A \\
		 RP_{l_1+1}+\ldots+RP_{l_2} + A \rightarrow B \nonumber \\
		 RP_{l_2+1}+\ldots+RP_{l_3} + B \rightarrow R^*, \nonumber
\end{eqnarray}
where $l_1$, $l_2$ and $l_3$ are the sizes of the three ribosomal protein groups and $l_1+l_2+l_3=54$.  

We find that as long as rRNA is not limiting, the two models lead to the same overall growth law as described by Eq. [1], under the following conditions; (i) If the limiting ribosomal proteins are the primary binders and we set $\tau_{SA} = \tau_{SA_1} + \tau_{SA_2} + \tau_{SA_3}$; (ii) If the secondary binders are limiting, and we set  $\tau_{SA}=\tau_{SA_2} + \tau_{SA_3}$; (iii) If the tertiary binders are limiting and we set $\tau_{SA}=\tau_{SA_3}$. This model could be readily generalized to accommodate actual assembly maps. See SI, section III for derivation of this result, and for treatment of the case where rRNA is limiting, and section II for an example of the distribution of assembly times for the 16S small ribosomal subunit, using Nomura's assembly map. 

 \subsection{Pipelined ribosome assembly --- derivation of the new growth law.} To derive Eq. \ref{EqOne}, consider a collection of ribosomes that are busy translating mRNA of various proteins and of ribosomal proteins.  Each ribosome, upon completing its current task, idle for a certain period $\tau_0$ after which with probability $\alpha$ it will synthesize a random ribosomal proteins, and with the complementary probability $\beta=1-\alpha$ will make another protein.
 
The ribosome lifetime is an important parameter whose manipulation affect the level of ribosomes in the cell relative to other proteins (and their lifetime), but its effect on the growth law is negligible as long as the life time is much larger than the doubling time \cite{PrionPaper}. Thus, with the purpose of deriving an equation for the growth rate in mind, we assume ribosome lifetime is infinite. 
 
In the first self-assembly model (Eq. \ref{SAreactionScheme}), if all $54$ ribosomal proteins exist in stoichiometry then self-assembly of a new ribosome is initiated. In the second self-assembly model, if the entire set of primary binding ribosomal proteins and their target rRNA exist, self-assembly is initiated. However, if the secondary binders and their target $A$ do not exist in stoichiometry, the second step in the assembly will be delayed, and similarly for the third stage (see Eq. \ref{SAreactionScheme2}) . 

Let $n_0(t)$ be the number of ribosome that enter the ``rest'' state at time $t$. Three processes contribute to $n_0(t)$. The first contribution is from the ribosomes that finished translating ribosomal proteins. The second contribution is from ribosomes that finished translating other proteins. The last contribution is from the ribosomes that were just finished being assembled. 

If all the durations are deterministic then the number of ribosomes that just finished translating ribosomal proteins is on average $\alpha n_0(t-\tau_0 - \frac{\tau_{RP}}{n})$ since $\alpha$ is the fraction of free ribosomes that upon exiting rest mode will be allocated for making ribosomal proteins, and the average duration of making a single ribosomal protein is $\frac{\tau_{RP}}{n}$. Similarly, the number of ribosomes that finished translating other proteins is $\beta n_0(t-\tau_0 - \tau_P)$, where $\beta=1-\alpha$, and $\tau_P$ is the average time to make a non-ribosomal protein.

The number of ribosomes that just finished being assembled at time $t$, $R^*(t)$ depends on the assembly model. Using the first assembly model (Eq. 3) yields $R^*(t)=\min_{i \in \{1,...,54\}} (n_1(t-\tau_{SA}),...,n_{54}(t-\tau_{SA}) )$, where $n_i(t)$ is the number of ribosomal proteins of type $i$ that completed being synthesized at time $t$.

On the other hand, the number of ribosomal proteins of type $i$ that finished being synthesized at time $t$ is given by $n_i(t)=\frac{\alpha}{n} n_0(t-\tau_0 - \frac{\tau_{RP}}{n})$, for all $i$. We thus obtain, 
\begin{eqnarray}\label{EqOneDer}
		n_0(t) = \alpha n_0(t- \tau_0 - \frac{\tau_{RP}}{n}) + \beta n_0(t- \tau_0 - \tau_{P}) + \nonumber \\ 
		\frac{\alpha}{n} n_0(t-\tau_0 -\frac{\tau_{RP}}{n} - \tau_{SA}).
\end{eqnarray}

This formula is valid under the assumption that $n_0(t) \gg 1$ such that there are many ribosomes concurrently working on all ribosomal protein and hence the completion time of all $54$ ribosomal proteins equals to $\frac{\tau_{RP}}{n}$ and the fraction allocated towards any particular ribosomal protein is $\frac{\alpha}{n}$. This homogeneity of translation rates and allocations is not essential but simplifies the analysis.
 
Equations \ref{EqOneDer} takes into account the pipelining in the reproduction process, where ribosomes are free to restart another round of translation of ribosomal proteins with probability $\alpha$ or other proteins with probability $\beta$, before the self-assembly process is finished. To obtain the growth law, we now assume that all the durations are random and independently distributed. Averaging over all possible durations, and taking the Laplace transform, we obtain 
\begin{eqnarray}\label{EqOneDer2}
		n_0(s) = n_0(s) P_0(s) \left(\alpha P(s) +\beta P_p(s) + \frac{\alpha}{n} P(s)P_{SA}(s) \right).
\end{eqnarray}
The average length of all non-ribosomal proteins in \textit{E. coli}'s genome is $2.3$ times larger than the average length of the ribosomal proteins. However, taking into account the average mass fraction of ribosomal proteins in the total proteome, which is measured to be $0.27$ \cite{hwa1,Bremer} at doubling times of 21.5 min, one finds that $P_p(s) =P(s)$, i.e. the actual load for making a generic protein is the same as the load of making a ribosomal protein. Inserting this to Eq. \ref{EqOneDer2} and dividing both sides by $n_0(s)$ we recover Eq. \ref{EqOne}.

\subsection{Serial ribosome assembly.} Consider next serial ribosome reproduction. In serial reproduction ribosomes cannot start a new round of reproduction before the current round is finished. To accommodate that, Eq. \ref{EqOneDer} has to be modified as follows:
\begin{eqnarray}\label{SerialLimit}
 n_0(t)=\alpha n_0(t-\tau_0-\frac{\tau_{RP}}{n} \boldsymbol{-\tau_{SA}})&+&\beta n_0(t- \tau_0-\tau_{P})\nonumber + \\ 
 \frac{\alpha}{n} n_0(t- \tau_0-\frac{\tau_{RP}}{n}-\tau_{SA})&&
\end{eqnarray}
where we emphasized the only change made in Eq. \ref{EqOneDer} --- the addition of a delay $\tau_{SA}$ for all the ribosomes that are involved in the process of making ribosomal protein subunits. This delay ensures that all the ribosomes that are involved in ribosomal protein synthesis will be able to continue to translate only after the new ribosome they were involved in making finish the assembly process. Hence, no new generation of ribosome is started before the previous generation is completed, and so, no overlapping reproduction cycles occurs in the serial model. The typical, tacit assumption is a limiting case of this serial model, whereas the assembly time is negligible compared to the synthesis of all the ribosomal proteins and henceforth dropped. 

Averaging over all durations in Eq. \ref{SerialLimit}, taking the Laplace transform as before and assuming $P_p(s) \approx P(s)$ we obtain $n_0(s) = n_0(s) P_{0}(s) P(s) P_{SA}(s)(\alpha+\frac{\alpha}{n})+\beta n_0(s)P_0(s)P(s)$. Dividing by $n_0(s)$ we obtain the serial growth law:
\begin{eqnarray}\label{SerialGrowthLaw}
 P_0(s)P(s)P_{SA}(s)(\frac{\alpha}{n}+\alpha)+\beta P_0(s)P(s)=1.
\end{eqnarray}
Both models --- pipelined self-replication (Eq. \ref{EqOne}) and serial self-replication (Eq. \ref{SerialGrowthLaw}), agree in the limit where the assembly time tends to zero $\tau_{SA} \rightarrow 0$.
\subsection{On the serial limit with zero idling and assembly time.} An important limiting case of Eq. \ref{EqOne} we now turn to consider is the case with zero idling and assembly durations. To get this limit we set $P_{SA}(s)=P_0(s)=1$ in Eq. \ref{EqOne}. The resulting equation is then $P(s)(1+\frac{\alpha}{n})=1$. Setting $P_{SA}(s)=P_0(s)=1$ in Eq. \ref{SerialGrowthLaw} leads to the same equation (see Fig. 2).
 
Lets assume that the duration for translating all ribosomal proteins is distributed exponentially with an average $\tau=\frac{\tau_{RP}}{n}$, i.e. all ribosomal proteins are produced in parallel (as before, $\tau_{RP}$ is the average duration to make all the ribosomal proteins, and $n=54$). Then the Laplace transform of the exponential distribution is given by $P(s)=\frac{1}{1+s \tau}$. This yields the growth law $\mu=\frac{\alpha}{\tau_{RP}}$, which is also the growth law for a serial assembly, where each ribosome makes all the ribosomal proteins.

Alternatively, we can assume that the time to make all $n$ ribosomal proteins is Deterministic by setting $P(s)=e^{-\frac{\tau_{RP}}{n} s}$. This yields the growth law $\mu=\frac{n}{\tau_{RP}} \ln \left(1+\frac{\alpha}{n} \right)  \approx \frac{\alpha}{\tau_{RP}}$, since $\alpha < 1$ and $n = 54 \gg 1$. For an extension to second order of this model, with interesting biological implication see \cite{shlomi}. Hence we conclude that in the absence of an assembly step, to first order, both exponential and deterministic distributions effectively yield the same growth rate, if the number of parallel processes is $n$ is large. 
\subsection{Comparing pipelined to serial self-replication.} 
\begin{figure}
\centering
\includegraphics[width=1 \columnwidth]{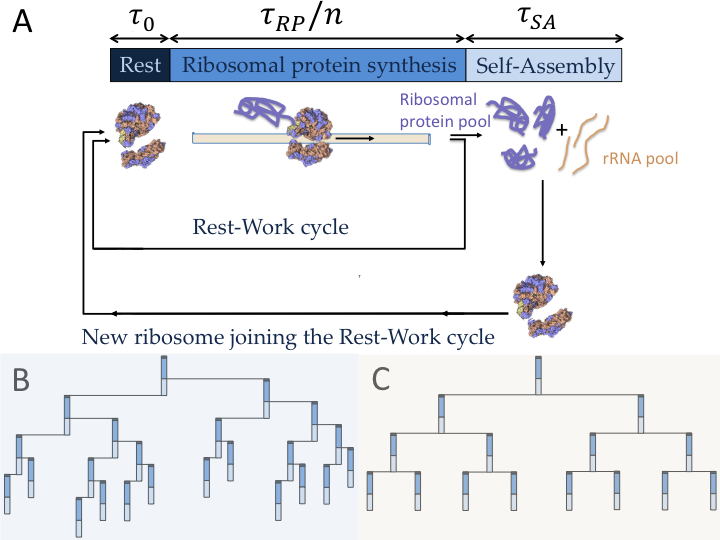} \\
\caption{\label{fig1}Model for pipelined self-reproduction of ribosomes. (A) Ribosomes translate ribosomal proteins probabilistically (with probability $\alpha$). After each translation they idle for a duration $\tau_0$, after which they return to translate proteins (ribosomal with probability $\alpha$ and other with the complementary probability). When ribosomal proteins accumulate in stoichiometry, the self-assembly process starts. While self-assembly proceeds, more ribosomal proteins are concurrently being synthesized. Upon completion of the self-assembly process a new ribosome is added to the pool of existing ribosome, and joins the rest-work cycle. Inset (B) shows the temporal structure of this process, with time running from up to bottom. Note the overlapping reproduction cycles, caused because ribosomes keep making ribosomal proteins with probability $\alpha$, concurrently with the assembly process. Inset (C) shows the corresponding serial process, where the old and the newly added ribosome, initiate a new reproduction round simultaneously.}
\end{figure}

In Fig. \ref{fig1}A we illustrate the pipelined self-reproduction model with deterministic variables. Ribosomes in the rest-work cycle translate ribosomal proteins or other proteins, then rest, then translate again. The ribosomal proteins enter into ``pools'' and whenever a full set exists, the self-assembly process of another ribosome is initiated. Self-assembly happens concurrently to the translation process, and, upon completion, the newly synthesized ribosome joins the rest-work cycle, starting from rest mode. Setting $\tau_{SA}=0$ yields a particular serial limit. Figure \ref{fig1}B illustrates the growing tree of ribosomes as a function of time --- advancing from top to bottom, and assuming for simplicity that $\beta=0$.  

Also noticeable in Fig. \ref{fig1}B, is the overlap between different generations, due to the fact that existing ribosomes start a new generation prior to the completion of the previous generation. For comparison,  in Fig. \ref{fig1}C presents the serial limit (also with $\beta=0$), whereas newly formed ribosomes and existing ones start a new round simultaneously. Evidently, for a given critical path duration for the formation of a single given ribosome $\tau_{RP}+n\tau_0+\tau_{SA}$ the serial limit has slower growth rate,  see also Fig. 2. 

\subsection{Self assembly step increase growth rate relative to the serial case for a given critical path duration.} To study the effect of overlapping reproduction cycles on the growth rate let us assume, hypothetically, that the parameters $\tau_{RP}$ and $\tau_{SA}$ are tunable, provided that their sum $\tau_{RP} + \tau_{SA}$ is kept constant. In addition we also keep $\tau_0$ constant, so overall, the critical path duration to make a single particular ribosome $\tau_{RP} + n \tau_0 + \tau_{SA}$ is assumed constant throughout the following discussion.
\begin{figure}
\centering
\includegraphics[width=1 \columnwidth]{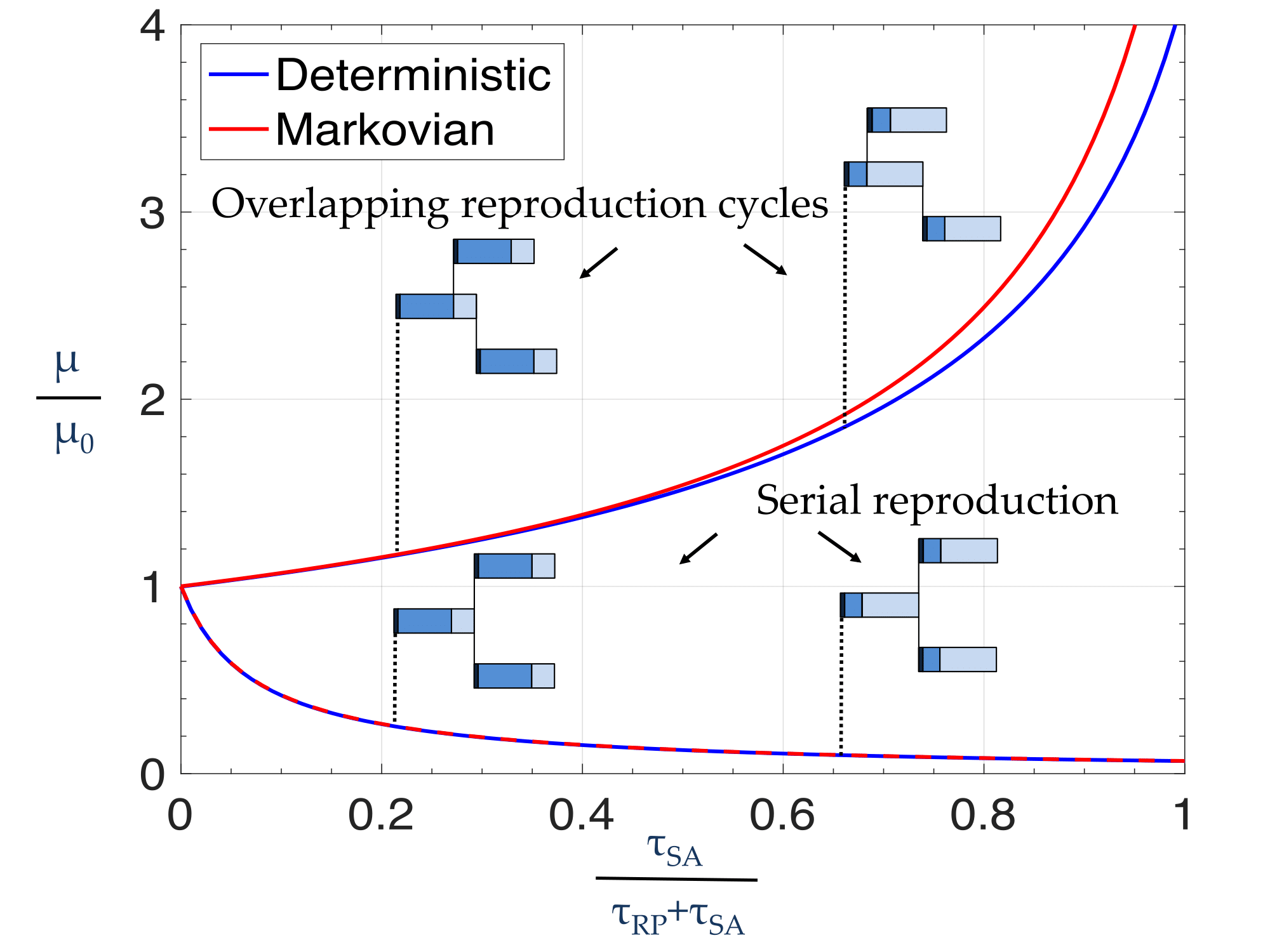} 
\caption{ \label{fig2}Comparison between the overlapping reproduction cycle model (Eq. \ref{EqOne}) and the serial model (Eq. \ref{SerialGrowthLaw}). Relative increase in growth rate for Markovian (red line) and deterministic (blue line) distributions, as a function of the percent of overlap between the self-assembly duration and the overall net duration of making a new ribosome $T=  \tau_{RP} + \tau_{SA}$ assumed constant. The parameter $\tau_{RP}$ is the ribosomal protein workload --- the time to make all the ribosomal proteins by a single ribosome. The nominal growth rate, $\mu_0=\frac{\alpha}{T + n \tau_0}$ represents the completely serial limit where the doubling time of a single ribosome is solely composed of the time to produce all its ribosomal proteins and the assembly time is set to zero. Using an allocation parameter $\alpha=28 \%$ as measured in \cite{RibosomeProfiling2} we calculated the growth rate using Eq. \ref{EqOne} as a function of the assembly time. A clear increase in growth rate is observed for the overlapping reproduction cycle model (as described by Eq. \ref{EqOne}), as well as a difference between completely Markovian (red upper line) and completely deterministic (blue lower line) limits. For comparison we plotted the relative growth rate for the serial growth law (as described by Eq. \ref{SerialGrowthLaw}), for which the relative growth rate decrease with the overlap parameter, and the difference between the Markovian and the deterministic distributions is negligible. To aid understanding we also present an inaccurate illustration of the structure of the temporal tree for overlap parameters $\frac{\tau_{SA}}{T}=0.21$ and $\frac{\tau_{SA}}{T}=0.66$ but with $\alpha=1$. Note that for $\frac{\tau_{SA}}{T}=50\%$ the increase in growth rate of the overlapping reproduction cycles model is by $50\%$.}
\end{figure}				

In Fig. \ref{fig2} we plot the overall growth rate as a function of the ratio between the self-assembly time to the constant net production time, $\frac{\tau_{SA}}{\tau_{RP}+\tau_{SA}}$. 

The monotonically increasing solid red line represent the case where the duration to make ribosomal proteins, the assembly duration, and the idling duration are all exponentially distributed (with an average duration of $\frac{\tau_{RP}}{n}$ and $\tau_{SA}$ and $\tau_0$ respectively). The monotonically increasing solid blue line represent the case where the durations to make ribosomal proteins, the assembly duration, and the idling duration are all deterministically distributed (with an exact duration of $\frac{\tau_{RP}}{n}$, $\tau_{SA}$ and $\tau_0$ respectively). The growth rate in the Markovian case is noticeably larger than in the deterministic case as seen in Fig. \ref{fig2}, in contrast to the serial limit. %This is due to the exponential dominance of all the lower than median random durations.

As seen in Fig. \ref{fig2}, as the overlap increases, the growth rate increases monotonically and non-linearly, due to the overlap in the reproduction cycles. Specifically, when the overlap parameter is $50\%$, the increase in growth rate relative to the serial limit, is $50\%$. To contrast, the monotonically decreasing blue and red dashed lines represent the Markovian (red) and deterministic (blue) serial limits, which are essentially identical. The reason for the decrease in the growth rate as a function of the overlap in the serial model is because as the assembly time increases, ribosomes translating ribosomal proteins are delayed for longer durations before they are allowed to start translating ribosomal proteins for the next generation. 

We emphasize the difference between the serial ribosome reproduction scenario with a zero assembly time and the overlapping ribosome reproduction scenario with an assembly time $\tau_{SA}$ that is equal to the synthesis time of all ribosomal proteins $\tau_{RP}$. When these two scenarios are compared with the same critical path duration for making a single ribosome $T_c=\tau_{RP}+n \tau_0 + \tau_{SA}$, the overlapping reproduction cycle scenario will have a growth rate that is $50\%$ larger than the serial ribosome reproduction scenario. This is in-spite of the fact that in both these scenarios at steady-state a new ribosome will be added whenever an existing ribosome completes translation of a single ribosomal protein. 

This is because in the overlapping reproduction scenario with $\tau_{SA}=\tau_{RP}$ the existing ribosomes will finish on average a new set of ribosomal protein subunits for the next generation, just when the previous generation completes the assembly of an earlier ribosome. In contrast, in the serial reproduction scenario with $\tau_{SA}=0$, all the ribosomes will finish on average one round, and only then will start another round. As we argue below, the latter scenario is what we predict for \textit{E. coli} growing in rich defined medium. 
\begin{figure}
\centering			
 \includegraphics[width=1 \columnwidth]{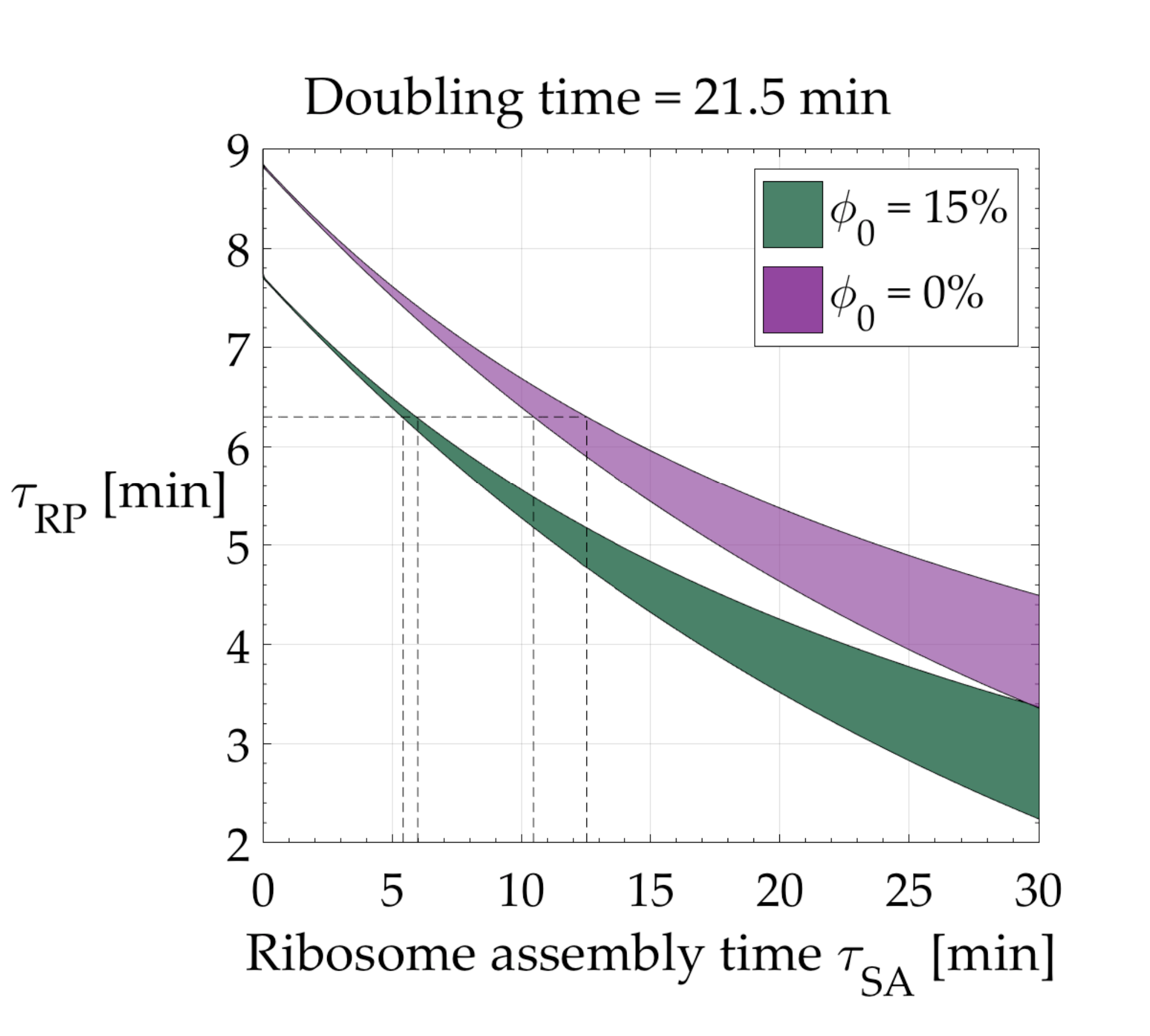}%{FigureOne18}
\caption{\label{fig4} Relation between the ribosome assembly time $\tau_{SA}$ and the ribosome protein workload $\tau_{RP}$ for a fixed doubling time of $21.5$ [min] as measured in \cite{RibosomeProfiling2}. Upper bound is set by a Markovian (Exponential) distribution. Lower bound is set by a deterministic distribution. Green area is calculated assuming the fraction of idling ribosome is zero $\phi_0=0$. Transparent purple area is calculated assuming that the fraction of idling ribosome is $\phi_0=15\%$. Dashed lines represent bounds on the in-vivo ribosome assembly time when the ribosome workload is $\tau_{RP}=6.3$ [min] which we deduce from an experimental measurements of average translation rate in a rich defined medium. The ribosome assembly time is at least $5.4$ min. If the fraction of idling ribosome is $\phi_0=10 \%$, the minimal ribosome assembly time is $\sim 7$ min.}
\end{figure}
\subsection{Calculating ribosome in-vivo assembly time.} We now turn to apply Eq. \ref{EqOne} for estimating the in-vio ribosome assembly time in \textit{E. coli} growing at $21$ min doubling time. Using Eq. \ref{EqOne} along with experimental measurements of ribosome cellular allocation fractions and the average translation speed, we can deduce upper and lower bounds on the in-vivo ribosome assembly time. This is done as follows. First, we use ribosome profiling data from \cite{RibosomeProfiling2} which reports that for growth in a rich defined (MOPS) medium at $37 ^o C$ the fraction of ribosomes translating ribosomal protein genes is $\alpha_{A} = 28\%$ (the doubling time in these conditions was $21.5 \pm 0.3$ min \cite{RibosomeProfiling2}). 

The fraction of idling ribosomes i.e. fraction of free ribosomes that are not attached to mRNA varies between different environments, but it is estimated to be around $10\% - 15\%$ in the same medium \cite{Bremer,RonsBook}. We found the value of $\tau_0$ that yields $\phi_0=15\%$ idling ribosomes to be $\tau_0= 0.02$ minutes. 

Next using measurements of in-vivo ribosome translation rates we calculated the ribosome workload --- the average duration for a single ribosome to translate all $54$ ribosomal protein subunits. The total number of amino-acids in all the ribosomal proteins (including the proteins that are present in duplicate) is $7249$ amino-acids. The average translation rate of a ribosome in rich defined medium is measured to be $19.2$ amino-acids per second, extracted from Fig. $3B$ in \cite{hwa2}, see also references therein. Thus, the ribosome workload is $\tau_{RP}=\frac{7249}{19.2}=377$ sec $=6.3$ min.

To calculate the in-vivo assembly time using these measurements we solved the following inverse problem. Given that the doubling time is $\frac{\ln 2}{\mu}=21.5$ min and that the global allocation parameter is $\alpha=0.28$ we numerically found all possible pairs of parameters $(\tau_{SA},\tau_{RP})$ that yield the same doubling time of $21.5$ min in Eq. \ref{EqOne}. For that purpose we had to define two extreme duration distributions for ribosomal protein synthesis and ribosome assembly. The first extreme distribution is a deterministic distribution, i.e. a distribution without any noise. 

The second extreme distribution is the distribution that has maximal noise given the average duration, which we derive by maximizing the entropy given the average. The maximal entropy distribution with a positive support (as durations cannot be negative) and a given average is  the exponential (Markovian) distribution \cite{dill}. In the absence of contrary evidence regarding the coefficient of variation being larger than one, any actual distribution of durations would have to lie somewhere in between these two extremes. We thus utilize these two distributions to set upper and lower bounds on the  assembly durations without worrying about the accuracy of these distributions as models for biological noises.
   
For ribosome workload of $\tau_{RP}=6.3$ min, and a ribosome idling fraction $\phi_0=0$, the assembly time in rich defined medium is found to lie between $10.5$ min (for exponentially distributed durations for translation and assembly) to $12.5$ min (for deterministic durations). When the fraction of idling ribosomes is $\phi_0=15\%$ the assembly time is estimated to be between $\tau_{SA}=5.4$ min using deterministic distributions, to $\tau_{SA}=6$ min, for exponential distributions. This is $\sim 3$ times higher than the estimate presented in \cite{RibosomeBook} --- two minutes in-vivo assembly time. Alternatively, setting the ribosome assembly time to $2$ minutes, the inferred ribosomal proteins translation rate reduces to $17$ aa/sec, which is different from the reported measurements that all lie in the range $18-21$ aa/sec, with the best fitted value for rich defined (MOPS) medium and doubling time of $21.5$ min being $19$ aa/sec \cite{hwa2,Bremer}. 
\begin{figure}
\centering
\includegraphics[width=1\columnwidth]{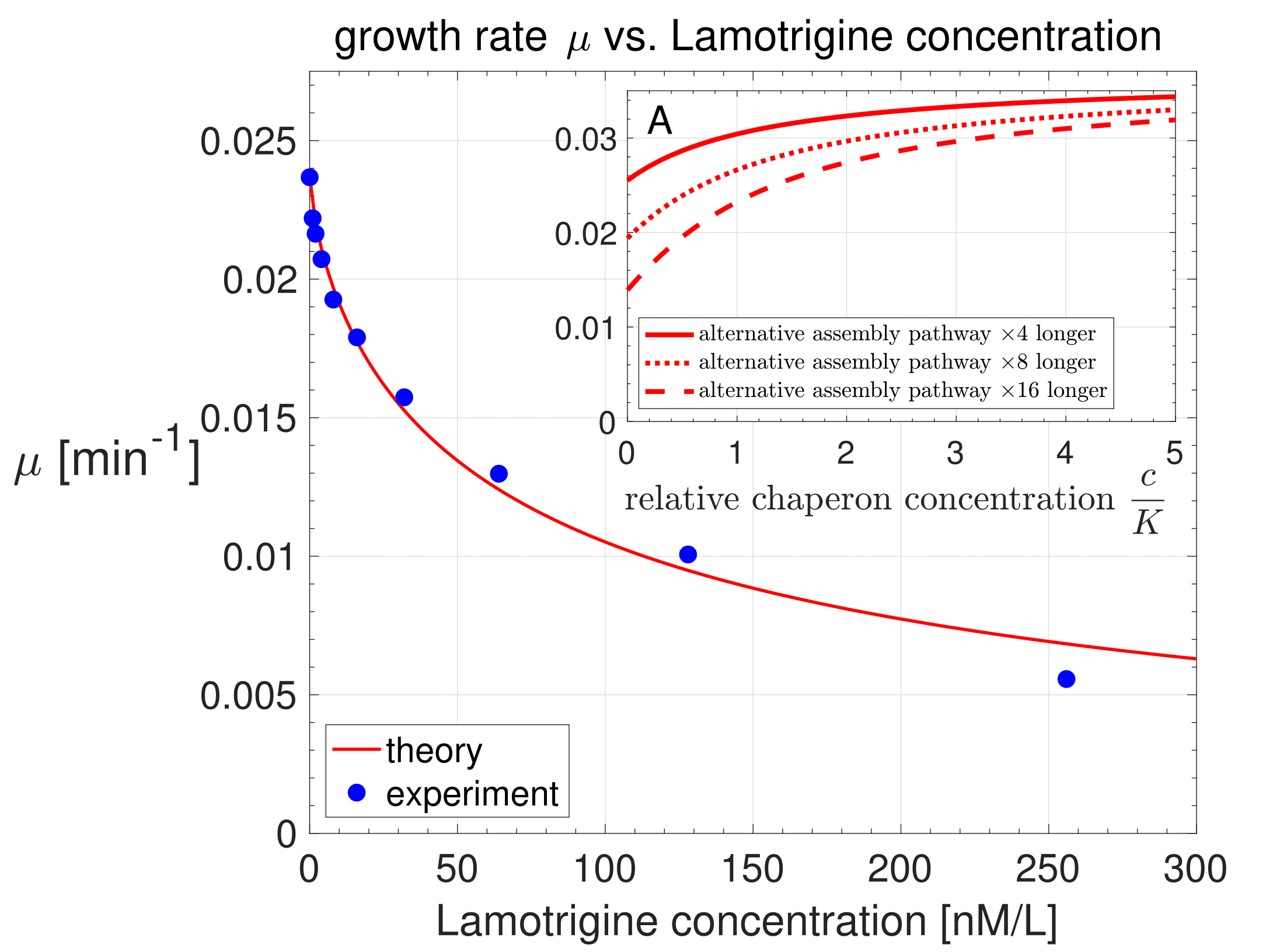}
\caption{\label{fig5}Growth rate as a function of Lamotrigine concentration. We calculated the growth rate as a function of Lamotrigine concentration assuming its biological mechanism halts assembly irreversibly. Blue circles are data points taken from \cite{Lamotrigine}. Red solid line is a fit using Eq. \ref{lam2} and \ref{lam1}. The two fitting parameters were the saturation parameter $K$ and Hill's factor $h$. Inset A shows the effect of limiting the expression of a ribosome assembly chaperon on the growth rate, assuming lack of chaperons may cause the traversed assembly pathway duration to be longer, e.g. because of the need to escape from kinetic traps that would have been avoided if chaperons were abundant. When chaperons are abundant ($\frac{c}{K} \gg 1$) the probability to reversibly fall to a kinetic trap is negligible and chaperons do not limit growth rate. On the other hand, when chaperons are deficient ($\frac{c}{K} \ll 1$) the probability to fall into a kinetic trap increases, which in turn cause an increased delays in the ribosome assembly time. We present three such delays; (i) Four times longer than the nominal (solid line); (ii) Eight times longer than nominal (dotted line); (iii) Sixteen times longer than nominal (dashed line). In the limit of a deep trap the delay tend to infinity and we recover the irreversible model.}
\end{figure}
 
\subsection{Modeling the effect of a drug that inhibits ribosome assembly.} In this section and the next we demonstrate that our modeling approach is versatile by applying it to quantitatively study phenotypes with faulty assembly of ribosomes. 

Recently, the drug Lamotrigine was reported to have a direct detrimental effect on ribosome assembly in live E. \textit{coli} \cite{Lamotrigine}. In the presence of Lamotrigine, E. \textit{coli} cells were observed to accumulate non-functional partially assembled ribosome complexes, which subsequently led to slower overall growth \cite{Lamotrigine}. We can model the effect of such a drug, by expanding our simplified model, adding a probability that the ribosome self-assembly process will fall into an irreversible trap, with a probability that depends on the Lamotrigine concentration. We use two parameters to characterize the probability of an assembly failure $p_{AF}$ as a function of the Lamotrigine concentration --- the saturation parameter $K$ and Hill's cooperativity factor $h$:
\begin{equation}\label{lam1}
p_{AF}(c)=\frac{K^h}{K^h + c^h}.
\end{equation}
The probability or an assembly failure goes to one when $c \gg K$, and tends to zero when $c \ll K$. 
The functional equation for the growth rate Eq. \ref{EqOne} is modified to read
\begin{equation}\label{lam2}
P_0(s)P(s)(\frac{\alpha}{n}P_{SA}(s)p_{AF}(c) +1)=1.
\end{equation}
Using Eq. \ref{lam2} to calculate the growth rate as a function of the Lamotrigine concentration, and the inferred assembly time from the previous section, we fitted the data from \cite{Lamotrigine} to obtain an estimate for $K$ and $h$. We find $K=52$ [nM/L] and $h=0.75$ indicating non-cooperative effect on the assembly. 

It would be of interest to biochemically measure $K$ and $h$ to ascertain our prediction and to elucidate the biochemical mechanism underlying the non-cooperative Hill factor. 
\subsection{Predicting the effect of limiting assembly chaperons.} In the previous section we assumed that the effect of Lamotrigine on the ribosome assembly process is irreversible. We also assumed that all the assembly chaperons are abundant. What if we want to model a situation in which the expression of a specified assembly chaperon is externally controlled? Clearly, if lack of certain assembly chaperons cause an effectively irreversible failure in the assembly process, we can utilize Eq. \ref{lam2} by adapting the $K$ and $h$ parameters to model the chaperon under consideration. However, the role of certain chaperons is to reduce the probability of falling intro a kinetic trap \cite{WilliamsonReview}. If the kinetic trap is ``shallow'' falling to it only slows down the assembly process. This cause two detrimental complementary effects; the average assembly duration increases and the coefficient of variation of the assembly duration increases. 

To model shortage of chaperons that cause such combined detrimental effect we introduce the Laplace transform for the modified distribution of assembly times $\hat{P}_{SA}(s)$ as a function of $p(c)$ --- the probability to choose a longer assembly pathway as a function of the concentration of the chaperon --- $c$. When the concentration of the chaperon protein far exceeds its saturation parameter $c \gg K$ the assembly time distribution is the previously used assembly time distribution $\hat{P}_{SA}(s)=P_{SA}(s)$. When $c \ll K$ the distribution shifts to a longer distribution which we assume for the sake of simplicity to be exponential, hence with Laplace transform that equals to $\frac{\lambda_c}{s+\lambda_c}$. We require that $\lambda_c> \frac{dP_{SA}(s)}{ds}{|_{s=0}}$ so that the alternative assembly route which the chaperon helps avoiding is longer than the nominal one on average. Hence,
\begin{equation}
\hat{P}_{SA}=\frac{c^h}{K^h+c^h }P_{SA}(s)+\frac{K^h}{K^h+c^h} \frac{\lambda_c}{s+\lambda_c}.
\end{equation}
It follows that the relation between the growth rate and the concentration of the chaperon will be given by,
\begin{equation}\label{chprn}
P_0(s)P(s)(\frac{\alpha}{n}\hat{P}_{SA}(s)+1)=1.
\end{equation}
In Fig. \ref{fig5}A we plot the relation between growth rate and the concentration of a chaperon protein that assist in assembly as described above, by solving Eq. \ref{chprn} . We show three types of kinetic traps; shallow (solid upper line); medium (dotted middle line) and deep (dashed lower line). From an evolutionary perspective we can expect the level of expression of ribosome assembly chaperon to be correlated with the ``severity'' of the kinetic trap it assist to mitigate. This can be tested by correlating their expression levels and detailed biophysical knowledge, which steadily accumulates, regarding the free energy landscape of ribosome assembly and the role of the chaperons in the assembly process.

	\end{article}

\end{document}